\documentclass[graybox]{svmult}

\usepackage{mathptmx}       
\usepackage{amsmath}
\usepackage{amsfonts}
\usepackage{amssymb}
\usepackage{url}
\usepackage{helvet}         
\usepackage{courier}        
\usepackage{type1cm}        

\usepackage{makeidx}         
\usepackage{graphicx}        
\usepackage{multicol}        
\usepackage[bottom]{footmisc}

\begin{document}

\title*{Resistive Switching Assisted by Noise}
\author{G. A. Patterson, P. I. Fierens and D. F. Grosz}
\institute{G. A. Patterson \at Facultad de Ciencias Exactas y Naturales, Universidad de Buenos Aires, Argentina \email{german@df.uba.ar}
\and P. I. Fierens \at Instituto Tecnol\'ogico de Buenos Aires (ITBA) and Consejo Nacional de Investigaciones Cient\'ificas y T\'ecnicas (CONICET), Argentina 
\and D. F. Grosz \at ITBA and CONICET, Argentina }

\maketitle

\abstract{We extend results by Stotland and Di Ventra \cite{stotland} on the phenomenon of resistive switching aided by noise. We further the analysis of the mechanism underlying the beneficial role of noise and study the EPIR (Electrical Pulse Induced Resistance) ratio dependence with noise power. In the case of internal noise we find an optimal range where the EPIR ratio is both maximized and independent of the preceding resistive state. However, when external noise is considered no beneficial effect is observed.}

\section{Introduction}
\label{sec:1}

In this work, we study a passive device called \textit{memristor} which was proposed by Chua \cite{chua} and has been modeled and extensively studied by, \textit{e}.\textit{g}., Strukov \textit{et al}. \cite{strukov1}. A memristor is a two-terminal device with the property that its resistance changes according with the electric charge that has flowed across it. The model in Ref. \cite{strukov1} reproduces, in a qualitative way, the resistive switching behavior of compounds such as TiO$ _{2} $. It consists of two resistors in series whose equivalent resistance depends on the dopant concentration, and it is described by a simple nonlinear dynamic equation.

The effect of resistive switching has been experimentally studied in many materials, such as simple oxides, and in more complex compounds such as manganites and cuprates, among others \cite{sawa, waser}. Results are usually reported as current-voltage hysteretic curves, or as the resulting resistance \textit{vs}. the externally applied field. One possible application of this phenomenon is in the design of non-volatile memories. Over the past decade extensive research was conducted in this direction by studying different properties, such as, retention times, temperature dependence, pulsing protocols, \textit{etc}. In order to quantify the switching performance, the EPIR ratio is defined as $ (R_{h} - R_{l})/R_{l} $, where $ R_{h} $ and $ R_{l} $ are the high and low non-volatile resistive states after input pulsing, respectively. A higher EPIR ratio is desirable in order to attain a larger contrast between resistive states.

\begin{figure}[h]
\begin{center}
\includegraphics[width=\textwidth]{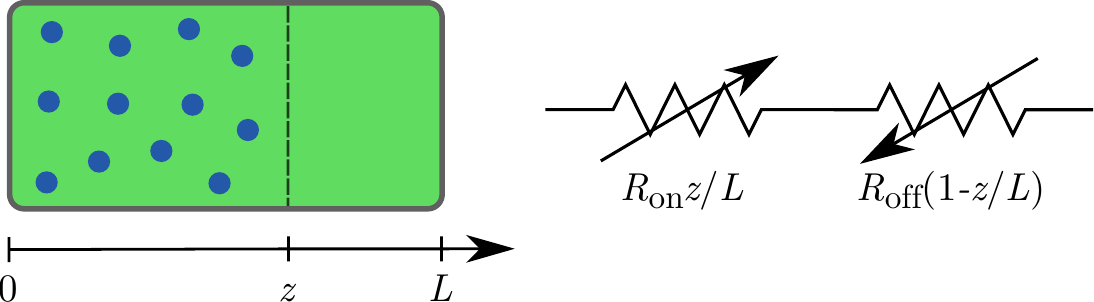}
\caption{The \textit{memristor} model consists of a sample divided into two doped/undoped regions with oxygen vacancies. The position $ z $ of the boundary between both regions determines the net resistance of the sample.}
\label{fig:scheme}
\end{center}
\end{figure}

In this work we extend results by Stotland and Di Ventra \cite{stotland} where the addition of white Gaussian noise to Strukov's equations was studied, showing that there is an optimal noise intensity that maximizes the contrast between high and low resistive states. They considered internal noise, inherent to the sample dynamics, and also suggested that the same results could be obtained if external noise was taken into account.

The paper is organized as follows: In Section \ref{sec:2} we review the model proposed by Strukov \textit{et al}. and introduce the stochastic model. In section \ref{sec:4} we show results of the stochastic model, and in Section \ref{sec:5} we present our conclusions.

\section{Resistive switching model and dynamics}
\label{sec:2}

Strukov \textit{et al}. \cite{strukov1} introduced a model of a memristor consisting of a sample of length $ L $ divided into two doped/undoped regions with oxygen vacancies. Fig. \ref{fig:scheme} shows a schematic of the sample model; where $z$ marks the boundary between both regions. The memristive system is described by

\begin{equation}
\begin{aligned}
V(t) & =  R(z)I(t) \ , \\
\frac{dz}{dt} & = f(z,I) \ ,
\label{ohm}
\end{aligned}
\end{equation}

\noindent where $ V(t) $ is the potential difference across the sample, $ I(t) $ is the current, and $ R $ is the sample resistance. Each region has a resistance that depends on the doping level. The net sample resistance is thus modeled as the resulting in-series resistance, and is given by

\begin{equation}
R(z) = R_{\mbox{{\small off}}}-(R_{\mbox{\small off}}-R_{\mbox{\small on}}) z/L \ ,
\label{memris}
\end{equation}

\noindent where $ R_{\mbox{\small off}} > R_{\mbox{\small on}} $ are the extreme possible values of resistance. The motion of the boundary is determined by the drift of vacancies produced by an external field and is given by (see, \textit{e}.\textit{g}., \cite{strukov1,blanc})

\begin{equation}
\frac{dz}{dt} = \frac{\mu R_{\mbox{\small on}} }{L}F(z)I(t) \ ,
\label{vel}
\end{equation}

\noindent where $ \mu $ is the average dopant mobility. The window function $F(z) = 1-\left(2z/L-1 \right)^{2}$ is introduced to account for experimentally observed nonlinearities and also enforces that $z$ remains within the interval $(0,L)$ \cite{strukov1, joglekar}.

Linearization of Eqs. \eqref{memris} around the two fixed points at $z = 0$ and $z = L$ leads to first order differential equations \cite{strogatz}

\begin{equation}
\begin{aligned}
\frac{dz}{dt} & \approx \frac{4 \mu R_{\mbox{\small on}} V(t)}{L^{2}R_{\mbox{\small off}}}z \ , \\
\frac{dz}{dt} & \approx -\frac{4 \mu V(t)}{L^{2}}(z-L) \ .
\label{lin}
\end{aligned}
\end{equation}

\noindent Note that, since $R_{\mbox{\small on}}/R_{\mbox{\small off}} < 1$, the boundary moves faster when it is close to the fixed point at $z = L$ than at $z=0$.

For convenience, we rewrite Eq. \eqref{vel} in a dimensionless form as

\begin{equation}
\frac{dx}{d\tau} = \frac{F(x)}{1-\Delta R x}v(\tau) \ , 
\label{system}
\end{equation}

\noindent where $ \Delta R = (R_{\mbox{\small off}} - R_{\mbox{\small on}})/R_{\mbox{\small off}} $, $ x  =  z/L $, $\tau  =  \mu A (1-\Delta R)t/L^{2} $, $ v(\tau) = V(t)/A $, and $ A $ is the amplitude of the external applied field in volts.

We considered the influence of both internal and external white Gaussian noise; internal noise affects the velocity of the internal state variable $ x $, while external noise is added to the input signal $ v $. 

\section{Results and discussion}
\label{sec:4}

\begin{figure}
\begin{center}
\includegraphics[width=80mm]{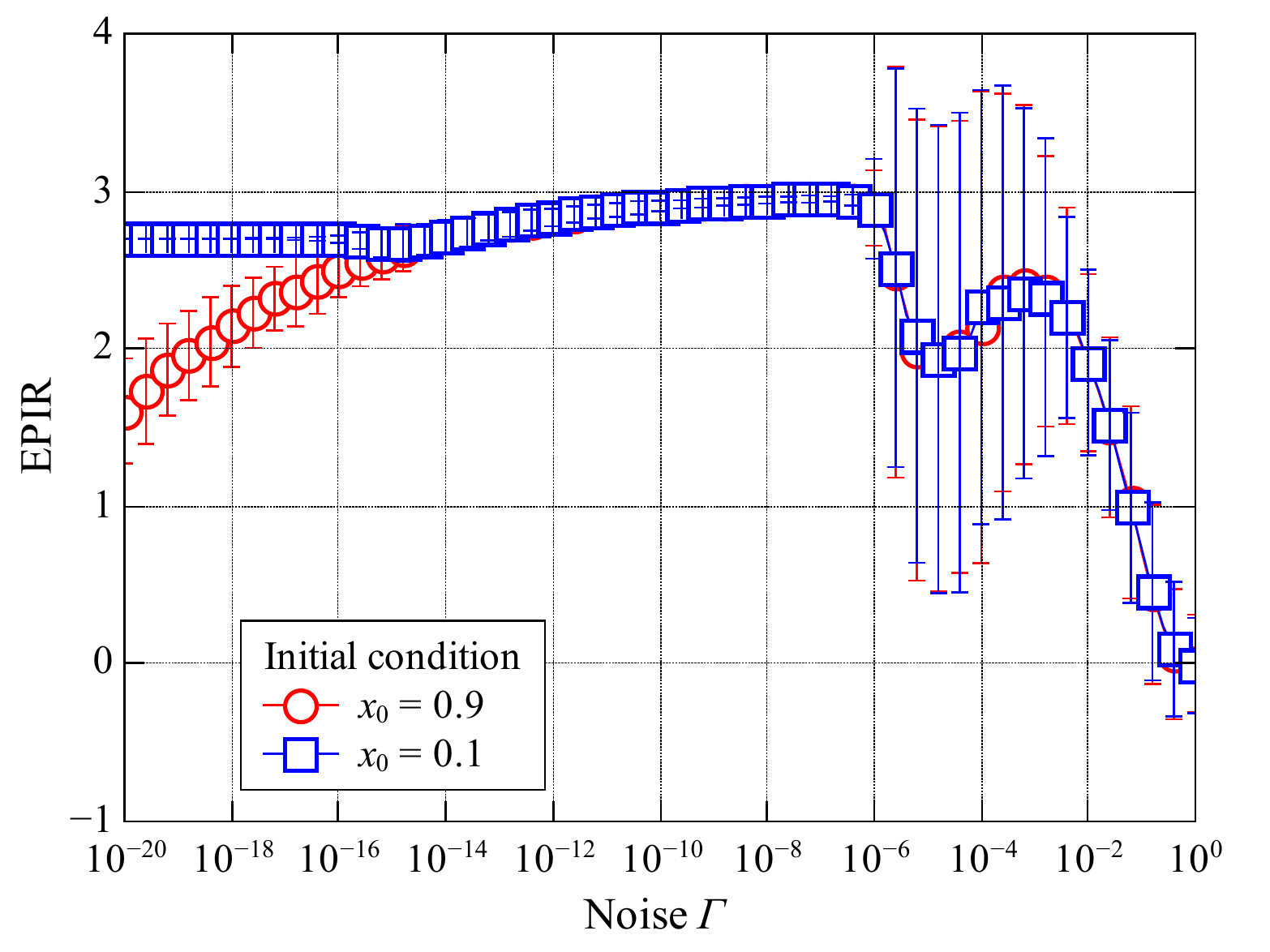}
\caption{EPIR ratio \textit{vs}. internal noise intensity averaged over 1000 realizations. Two maxima  at $ \Gamma \approx 10^{-7},  10^{-3} $ are observed. Error bars are taken as one standard deviation. Note the erratic behavior for $ \Gamma \gtrsim  10^{-6} $.}
\label{fig:int_epir_p}
\end{center}
\end{figure}

The externally applied field consisted of a sequence of pulses $+1$ $\rightarrow$ $0$  $\rightarrow$ $-1$  $\rightarrow$ $0$ repeated 5 times. The pulsewidth was 2, $ \Delta R = 0.75 $ and the resistance was computed during the last repetition. We solved Eq. \eqref{system} for the time evolution of the resistance for each noise realization and intensity. In Fig. \ref{fig:int_epir_p} the EPIR ratio is shown as a function of the noise intensity $ \Gamma $. The initial conditions are $ x_{0} = 0.9 $ (circles) and $ x_{0} = 0.1 $ (squares). The same behavior is observed for both initial conditions, namely the EPIR ratio increases with noise until it reaches a maximum at $\Gamma \approx 10^{-7}$. A second peak is observed at $ \Gamma \approx 10^{-3} $, but with a large standard deviation caused by the strong noise at that point. It is important to note that from Fig. \ref{fig:int_epir_p} the EPIR ratio is nearly independent of the initial condition and that standard deviations are small for $ \Gamma \in \left[ 10^{-15}, 10^{-7} \right] $.

One of the effects of noise is to help the boundary between the doped/undoped regions in the sample to escape from $x=0,1$. Eqs. \eqref{lin} suggest that the first maximum is reached when noise helps the boundary to escape from the border at $ x = 1 $. Since the boundary moves slower at $ x = 0 $, a stronger noise intensity is required to detach it, thus leading to a second maximum.

\begin{figure}
\begin{center}
\includegraphics[width=80mm]{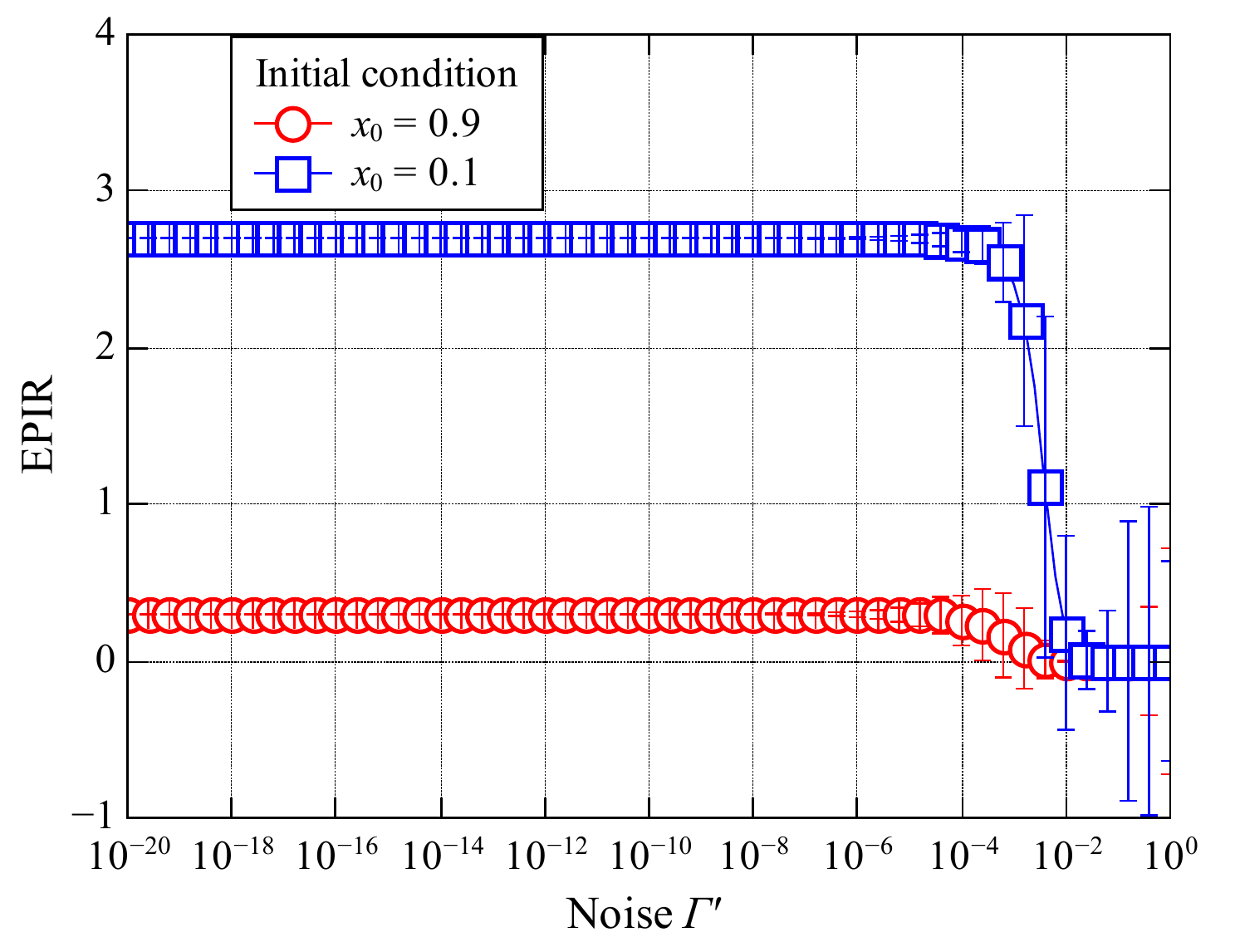}
\caption{EPIR ratio \textit{vs}. external noise intensity averaged over 1000 realizations. No maxima are observed and the EPIR ratio degrades for noise intensities $ \Gamma' \gtrsim 10^{-5} $.}
\label{fig:ext_epir_p}
\end{center}
\end{figure}

When external noise was considered, we did not find a noise intensity that maximizes the EPIR ratio. In Fig. \ref{fig:ext_epir_p} results are shown for the same set of parameters and initial conditions used before. In this case the window function effectively reduces the contribution of noise close to the sample borders. However, when the noise intensity is strong enough to counteract the window effect, the motion of the boundary is erratic, leading to a large standard deviation, and the EPIR ratio depends strongly on the initial condition.

\begin{figure}
\begin{center}
\includegraphics[width=80mm]{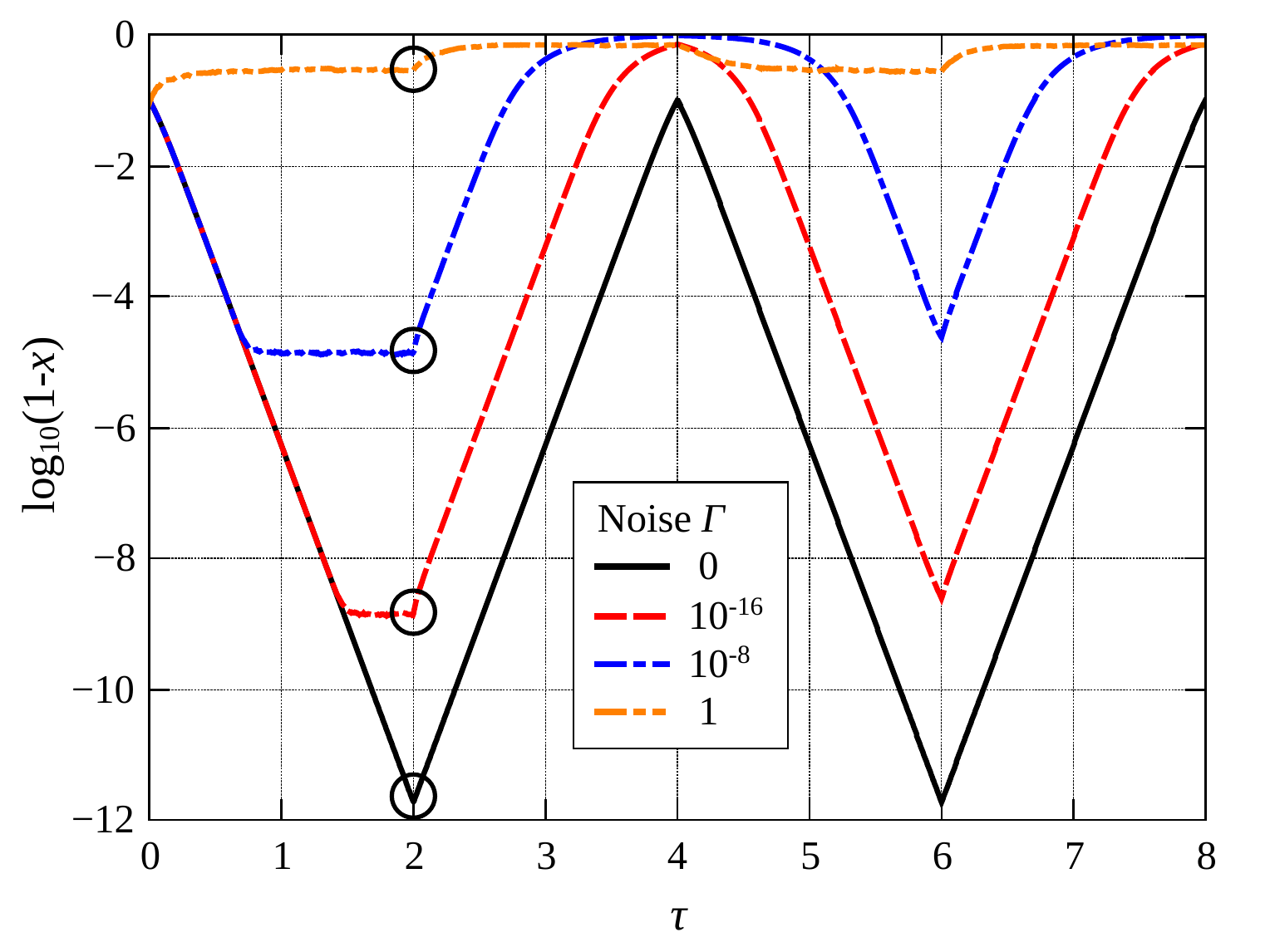}
\caption{Average boundary time evolution \textit{vs}. noise intensity. Positions after the first pulse (marked with circles) are noise dependent, leading to a new dynamic condition.}
\label{fig:pos_temp}
\end{center}
\end{figure}

In Fig. \ref{fig:pos_temp} the time evolution of the boundary is shown \textit{vs}. noise intensity. In this case, the applied sequence was $+1$ $\rightarrow$ $-1$  $\rightarrow$ $+1$  $\rightarrow$ $-1$, the pulsewidth was 2, and the initial condition was $ x = 0.9 $. We found a relation between the position reached by the boundary at the end of the first pulse, $x_n$ -- marked with circles in Fig. \ref{fig:pos_temp} -- and noise intensity: $1-x_{n} \propto \sqrt{\Gamma}$.

\begin{figure}
\begin{center}
\includegraphics[width=80mm]{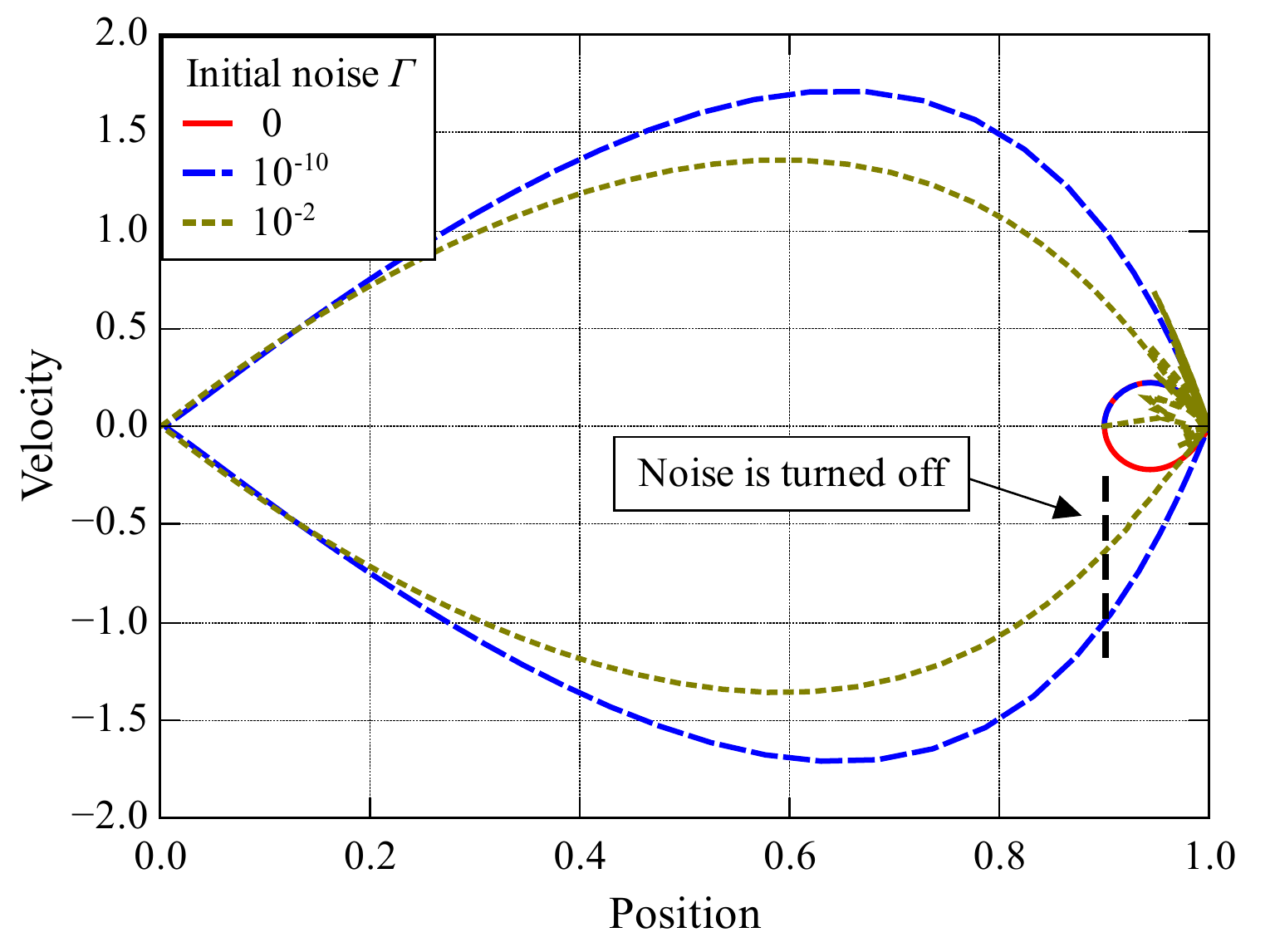}
\caption{Phase space trajectories for three different initial noise intensities. When the trajectories cross $ x = 0.9 $ after the first voltage reversal the noise is turned off.}
\label{fig:int_ruido_off}
\end{center}
\end{figure}

Phase space trajectories are shown in Fig. \ref{fig:int_ruido_off} for the noiseless and internal noise ($ \Gamma  = 10^{-10}$ and $ 10^{-2} $) cases. The applied signal is $ v(\tau) = \sin{\left( \omega \tau \right)} $, the initial position is $ x_{0}=0.9 $, and the angular frequency $ \omega  = 1$. In the noiseless case\footnote{The deterministic position of the boundary as a function of time can be obtained by direct integration of Eq. \eqref{system} and shown to be periodic with period $2/\omega$.}, the trajectory is a small closed loop. The boundary starts the motion from $ x_{0} $ towards $ x = 1 $; at time $ \tau = \pi/\omega $ the velocity is zero and the applied field is reversed, then it increases and the boundary moves in the opposite direction until it reaches the initial state at $ \tau = 2 \pi/\omega $. Larger loops are obtained when  considering a moderately strong internal noise in the system; in this case, at time $ \tau = \pi/\omega $ the velocity differs from zero as can be seen in Fig. \ref{fig:pos_temp}. In fact, if noise is turned off, for instance at $ (\tau_{\mbox{\small{off}}}, x_{\mbox{\small{off}}}=0.9) $ as shown in Fig. \ref{fig:int_ruido_off}, the motion will remain in the orbit determined by the new initial conditions $(\tau_{\mbox{\small{off}}},x_{\mbox{\small{off}}})$ and deterministic Eq. \eqref{system}. This way we see that noise has an important effect only when the boundary is close to the sample borders. At these points, in the case of internal noise, the velocity is strongly dependent on the noise intensity and leads to different trajectories; in the case of external noise, the window function reduces the influence of noise and no significant effect is observed.

\section{Conclusions}
\label{sec:5}

In summary, we presented results on the phenomenon of resistive switching under the influence of noise. On the one hand, we found that, when internal noise is considered, there is a range of noise intensities where the EPIR ratio is maximized and independent of the initial conditions. On the other hand, external noise only has the effect of degrading the EPIR ratio since strong intensities are needed to counteract the window-function effect. Moreover, we showed that, when the boundary is close to the borders of the sample, its velocity is strongly influenced by noise. We believe these results may be of relevance in systems where the large scale of electronic integration renders the effect of noise unavoidable.

\acknowledgement We gratefully acknowledge financial support from ANPCyT under project PICT-2010 \# 121.

\end{document}